\documentclass[
final            
]{aipproc}

\usepackage{fancyhdr}

\layoutstyle{6x9}

\begin{document}

\title{Nucleon structure from 2+1 flavor domain wall QCD at nearly physical pion mass}


\author{Shigemi Ohta for RBC and UKQCD Collaborations}{
  address={Institute of Particle and Nuclear Studies, KEK, Tsukuba, Ibaraki 305-0801, Japan},
  altaddress={Department of Particle and Nuclear Physics, SOKENDAI, Hayama, Kanagawa 240-0193, Japan},
  altaddress={RIKEN-BNL Research Center, Brookhaven National Laboratory, Upton, NY 11973, USA }
}

\begin{abstract}
The RBC and UKQCD collaborations have been investigating hadron physics  in numerical lattice quantum chromodynamics (QCD) with (2+1) flavors of dynamical domain wall fermions (DWF) quarks that preserves continuum-like chiral and flavor symmetries. 
The strange quark mass is adjusted to physical value via reweighting and degenerate up and down quark masses are set as light as possible.
In a recent study of nucleon structure we found a strong dependence on pion mass and lattice spatial extent in isovector axialvector-current form factors.
This is likely the first credible evidence for the pion cloud surrounding nucleon.
Here we report the status of nucleon structure calculations with a new (2+1)-flavor dynamical DWF ensembles with much lighter pion mass of 180 and 250 MeV and a much larger lattice spatial exent of 4.6 fm.
A combination of the Iwasaki and dislocation-suppressing-determinant-ratio (I+DSDR) gauge action and DWF fermion action allows us to generate these ensembles at cutoff of about 1.4 GeV while keeping the residual breaking of chiral symmetry sufficiently small.
Nucleon source Gaussian smearing has been optimized.
Preliminary nucleon mass estimates are 0.98 and 1.05 GeV.
\end{abstract}

\maketitle

\thispagestyle{fancy}
\lhead{Talk at Tropical QCD 2010, Cairns, QLD, September 27-October 1, 2010}
\rhead{KEK-TH-1438, RBRC-884}
\renewcommand{\headrulewidth}{0pt}
\cfoot{}

\section{Discovery of pion cloud?}

In a recent numerical lattice-QCD study \cite{Yamazaki:2008py,Yamazaki:2009zq} the RIKEN-BNL-Columbia (RBC) and UKQCD collaborations reported a strong dependence on pion mass, \(m_\pi\), and lattice spatial extent, \(L\), in isovector axialvector-current form factors of nucleon.
The dependence is most evident in the isovector axial charge, \(g_A\),
\begin{figure}[tb]
\includegraphics[width=.7\linewidth]{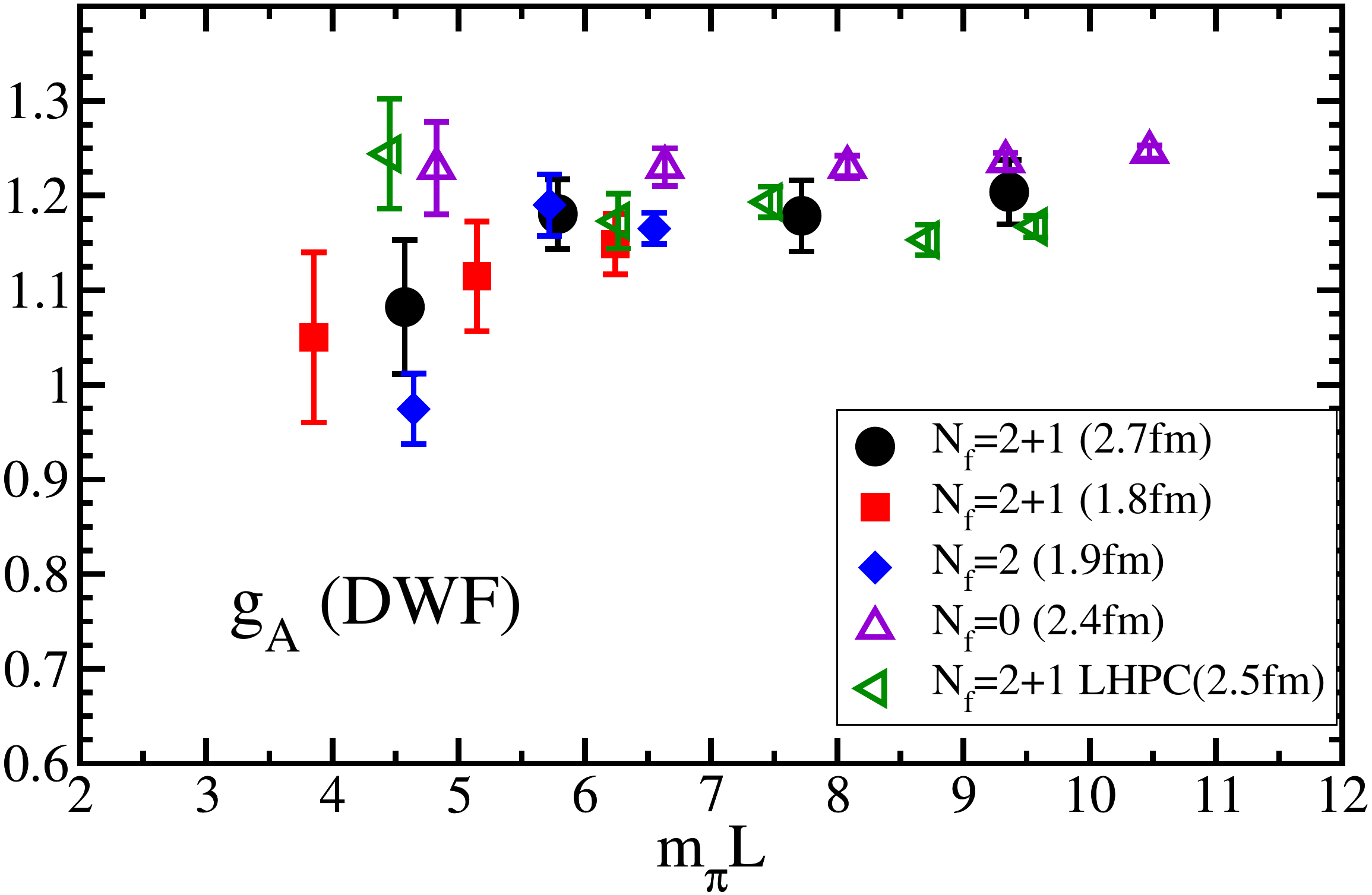}
\caption{Nucleon isovector axial charge, \(g_A\), calculated in numerical lattice QCD with DWF quarks plotted against a dimensionless variable, \(m_\pi L\), the product of calculate pion mass, \(m_\pi\), and lattice spatial extent, \(L\).
Results from RBC/UKQCD unitary (2+1)-flavor dynamical quark calculations show scaling in this variable with strong dependence.
In copmarison, non-unitary DWF calculations with either quenched or MILC ensembles show only weak, if at all, dependence.}
\label{fig:gAmpiL}
\end{figure}
as summarized in Fig.\ \ref{fig:gAmpiL}.
The dependence is much stronger than had been observed in non-unitary calculations with DWF quarks using either quenched \cite{Sasaki:2003jh} or rooted-staggered fermion \cite{Edwards:2005ym} ensembles.
As the dependence appears to scale with a dimensionless quantity, \(m_\pi L\), the product of calculated pion mass, \(m_\pi\), and the lattice spatial extent, \(L\), a likely explanation for the dependence is that significant part of the nucleon isovector axialvector current is carried by the expected, but never seen, ``pion cloud'' surrounding the nucleon.
If confirmed, this calculation may be the first concrete evidence of such a pion cloud.

Indeed similar strong dependence on pion mass and lattice spatial extent is seen in other axialvector-current form factors but not in the conserved vector-current ones \cite{Yamazaki:2009zq} or in low moments of the structure functions \cite{Aoki:2010xg}, supporting the pion-cloud interpretation.
However, since the calculations so far have only been carried out at single lattice cut off, \(a^{-1}\), of about 1.7 GeV, two lattice spatial extent, \(L\), of about 1.8 and 2.7 fm, and relatively heavy pion mass down to 330 MeV, it seems premature to conclude that the dependence is caused by the pion cloud and so we have discovered it.
This provides a good motivation for the new study being reported here with pion mass as light as 180 MeV and lattice spatial extent of about 4.6 fm.

\section{RBC/UKQCD (2+1)-flavor dynamical DWF lattice QCD}

For the past several years the RBC and UKQCD collaborations jointly generated two sets of (2+1)-flavor dynamical domain-wall fermions (DWF) lattice QCD ensembles at lattice cut off, \(a^{-1}\), of about 1.7 \cite{Allton:2008pn} and 2.2 GeV \cite{Aoki:2010dy}.
We adjust the strange quark mass to the physical value by reweighting \cite{Jung:2010jt,Aoki:2010dy}.
The degenerate up and down quark masses are set at several values, as light as possible, in the range corresponding to pion mass from about 420 MeV to 290 MeV.
In contrast to most other lattice fermion methods, the DWF formulation provides good control of chiral and flavor symmetries and non-perturbative renormalization.
This allows us to use meson observables such as pion and kaon mass, \(m_\pi\) and \(m_K\), and decay constants, \(f_\pi\) and \(f_K\), obtained from these ensembles, for a detailed study of meson physics in combined chiral and continuum limit \cite{Aoki:2010dy}: we first use pion, kaon, and \(\Omega\)-baryon mass to set the physical quark mass and lattice scale and then obtain predictions for other observables such as pion and kaon decay constants with a few \% accuracy.

At this level of high accuracy our predictions is now not limited by our statistics but by poor applicability of chiral perturbation or other chiral extrapolation ansatz from relatively heavy pion mass we use.
Thus the RBC and UKQCD collaborations are now jointly generating a third set of (2+1)-flavor dynamical domain-wall fermions (DWF) lattice QCD ensembles at lower pion mass of 180 and 250 MeV \cite{Jung:2010jt,Mawhinney:LAT2010}.
This is made possible by using a new combination for gauge action of Iwasaki and dislocation-suppressing-determinant ratio (DSDR) \cite{Vranas:1999rz,Vranas:2006zk,
Renfrew:2009wu} that allows a lower cut off of about 1.4 GeV while keeping the residual breaking of chiral symmetry sufficiently small.
The lower lattice cut off also allows large spatial extent of about 4.6 fm.

Such a study, however, requires good understanding of chiral and finite-size corrections to the observables which are unfortunately missing for baryons in general.
For this reason we are not yet attempting a similar combined chiral and continuum limit study for baryons at this moment.
Rather, we are refining our fixed-cutoff study of nucleon to improve our understanding of its chiral behavior.
We have published our findings in nucleon structure from the 1.7 GeV ensembles in three publications, Refs.\ \cite{Yamazaki:2008py,Yamazaki:2009zq,Aoki:2010xg}, which we summarize in the following.
As we discovered unacceptably large and distorting dependence on the scaling parameter \(m_\pi L\) of axialvector current form factors, especially at \(m_\pi L = 4.5\), we did not attempt a similar study for our 2.2 GeV ensembles to save our computing resources, and skipped to the new I+DSDR DWF ensembles with lighter pion mass and larger lattice spatial extent: the \(m_\pi L\) of these ensembles are respectively at 4.2 and 5.8 and should allow us to better understand the observed dependence.

\section{Nucleon structure at 1.7 GeV}

I here summarize the nucleon structure calculations \cite{Yamazaki:2008py,Yamazaki:2009zq,Aoki:2010xg} by RBC and UKQCD collborations using the 1.7-GeV dynamical (2+1)-DWF ensembles \cite{Allton:2008pn}.

In an earlier study with two dynamical DWF flavors \cite{Lin:2008uz}, we identified an important source of systematic error in lattice-QCD numerical calculation of nucleon structure, namely excited-state contamination.
\begin{figure}[b]
\includegraphics[width=.5\linewidth,clip]{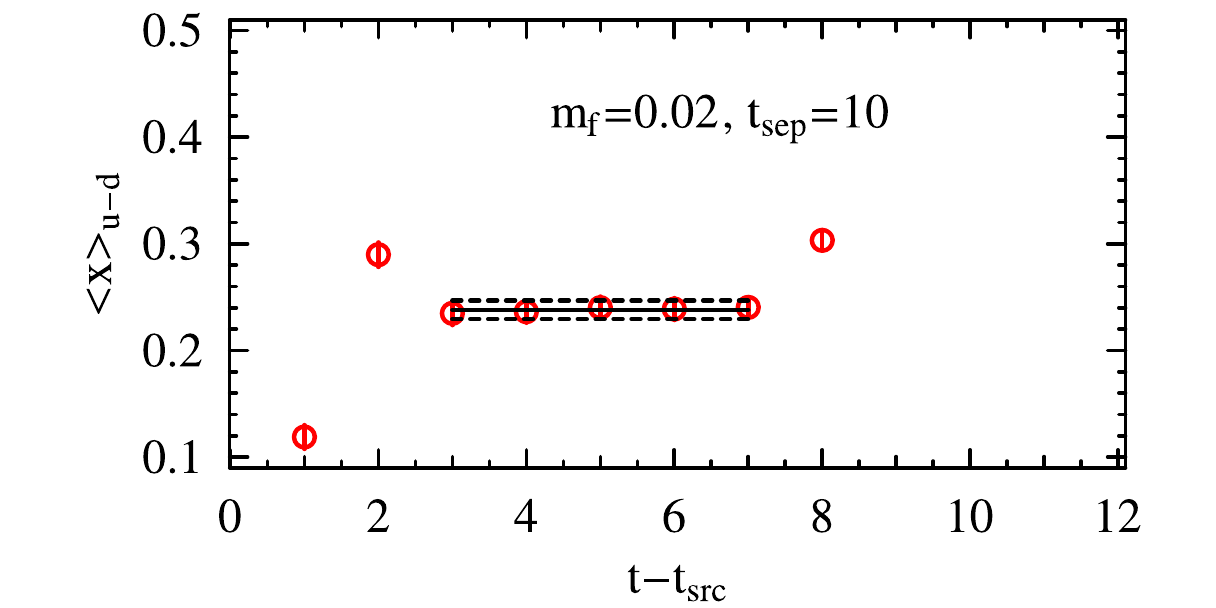}
\includegraphics[width=.5\linewidth,clip]{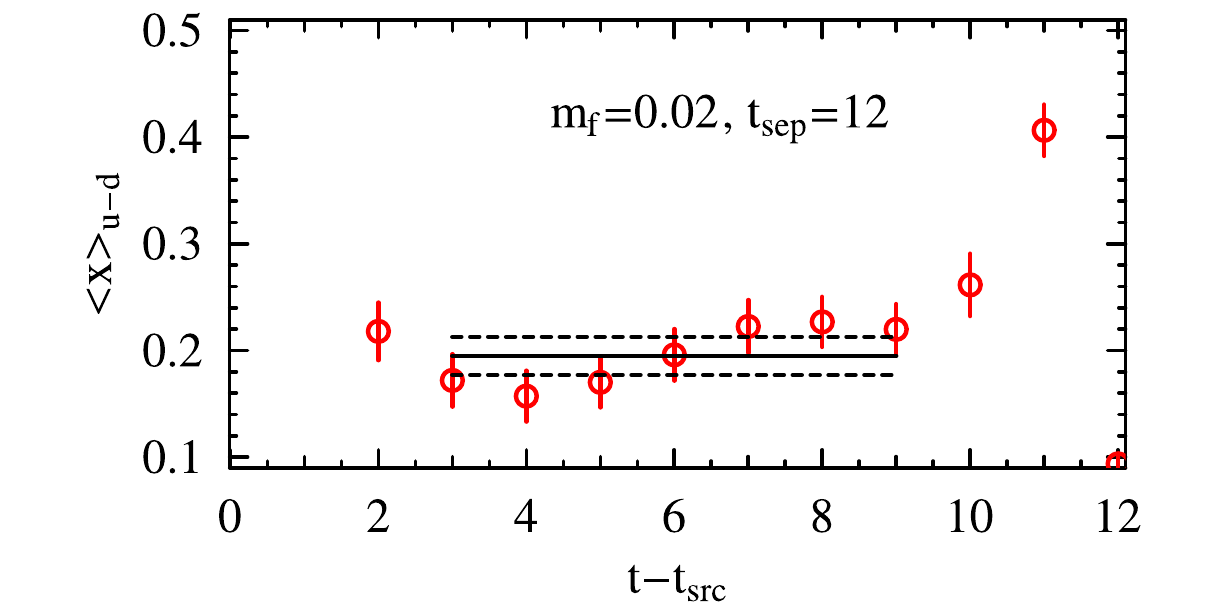}
\caption{In 2-flavor dynamical DWF study  \cite{Lin:2008uz}, systematic difference was identified arising from excited-state contamination.
Source smearing and source-sink separation must be optimized to filter out the excited state while maintaining reasonable statistical signal.
}
\label{fig:excitedcontamination}
\end{figure}
As the amount of contamination varies depending on the shape of source smearing, we need to optimize the combination of source smearing and source-sink separation in order to filter out the excited-state contamination while maintaining reasonable statistical signal.
If we choose too long a separation then even the ground state decays and no signal is obtained.
For our choice of Gaussian source smearing width of 7 lattice units, a source-sink separation of about 12 was optimal.

After making this important adjustment we discovered the isovector axial charge strongly depends on pion mass and lattice size, as was discussed earlier with Fig.\ \ref{fig:gAmpiL}.
The dependence seems to scale in a single parameter, the product \(x=m_\pi L\) of the pion mass \(m_\pi\) and lattice size \(L\).
Though our dynamic range in the scaling parameter is rather narrow to distinguish various ansatze on functional form, \(f(x)\), of this scaling, and so cannot yet clarify if this is indeed the pion cloud, fitting to various forms such as \(x^{-3}\) (inverse volume) or \(e^{-x}/\sqrt{x}\) (pion cloud) is possible, and results in an estimate of \(g_A=1.20(6)_{\rm stat}(4)_{\rm syst}\).
This strong dependence on \(m_\pi L\) is observed also in other axialvector-current form factors.
In order to drive the systematic error arising from this dependence below 1 \%, we would need \(m_\pi L\) of 6 to 8, or \(L\) of about 5 fm for \(m_\pi\) of 300 MeV and 10 fm for physical pion.

\begin{figure}[tb]
\includegraphics[width=0.33333\linewidth,clip]{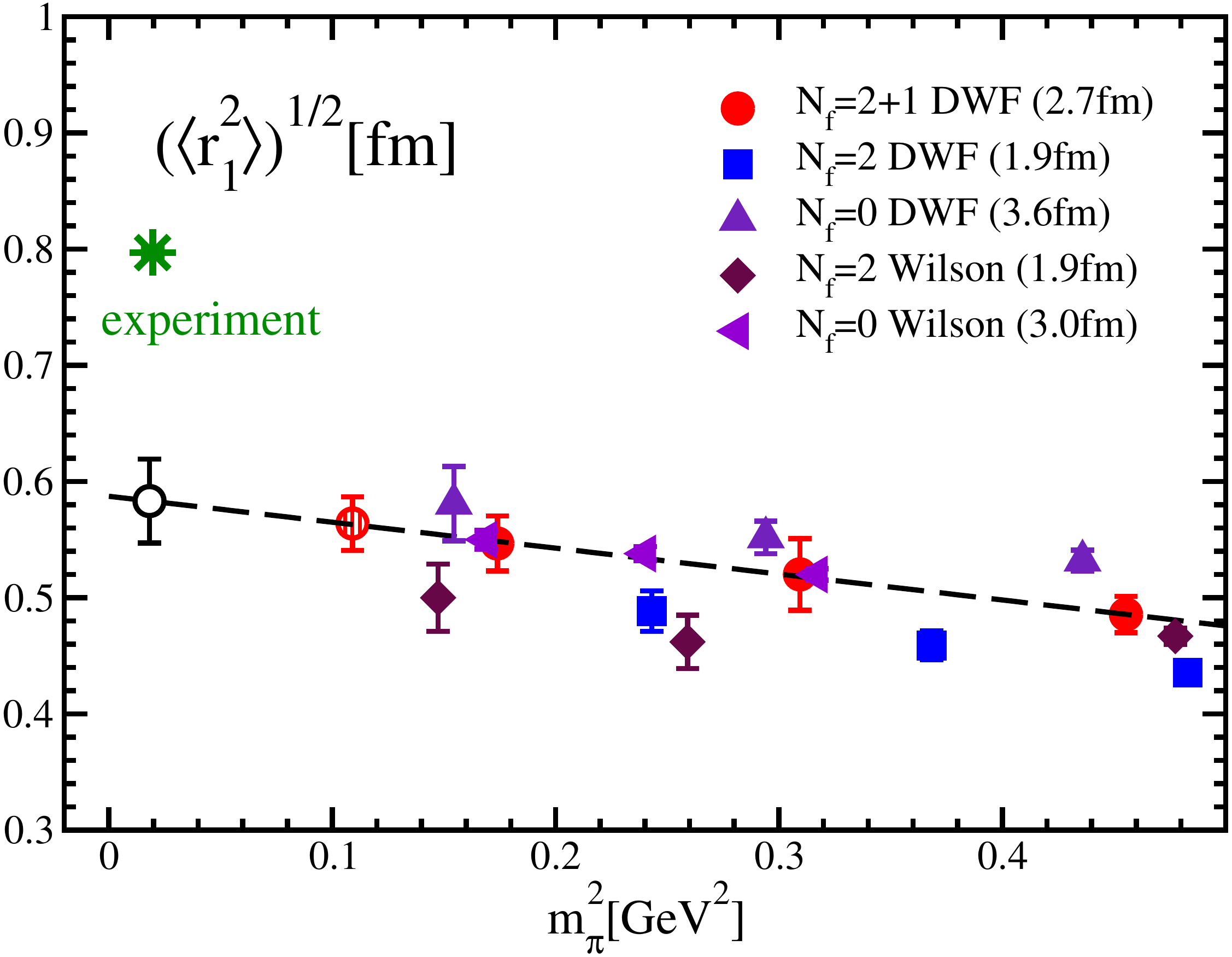}
\includegraphics[width=0.33333\linewidth,clip]{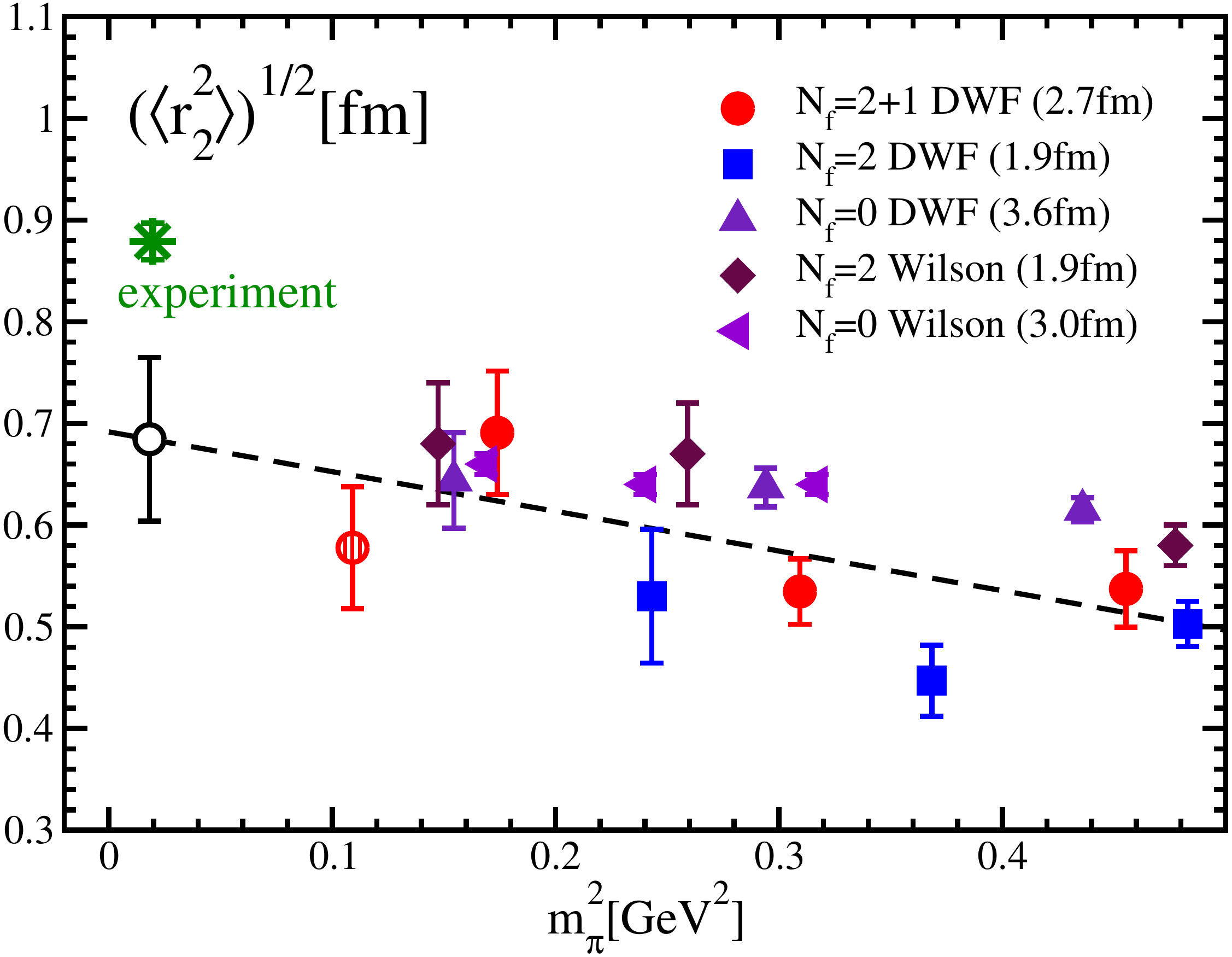}
\includegraphics[width=0.33333\linewidth,clip]{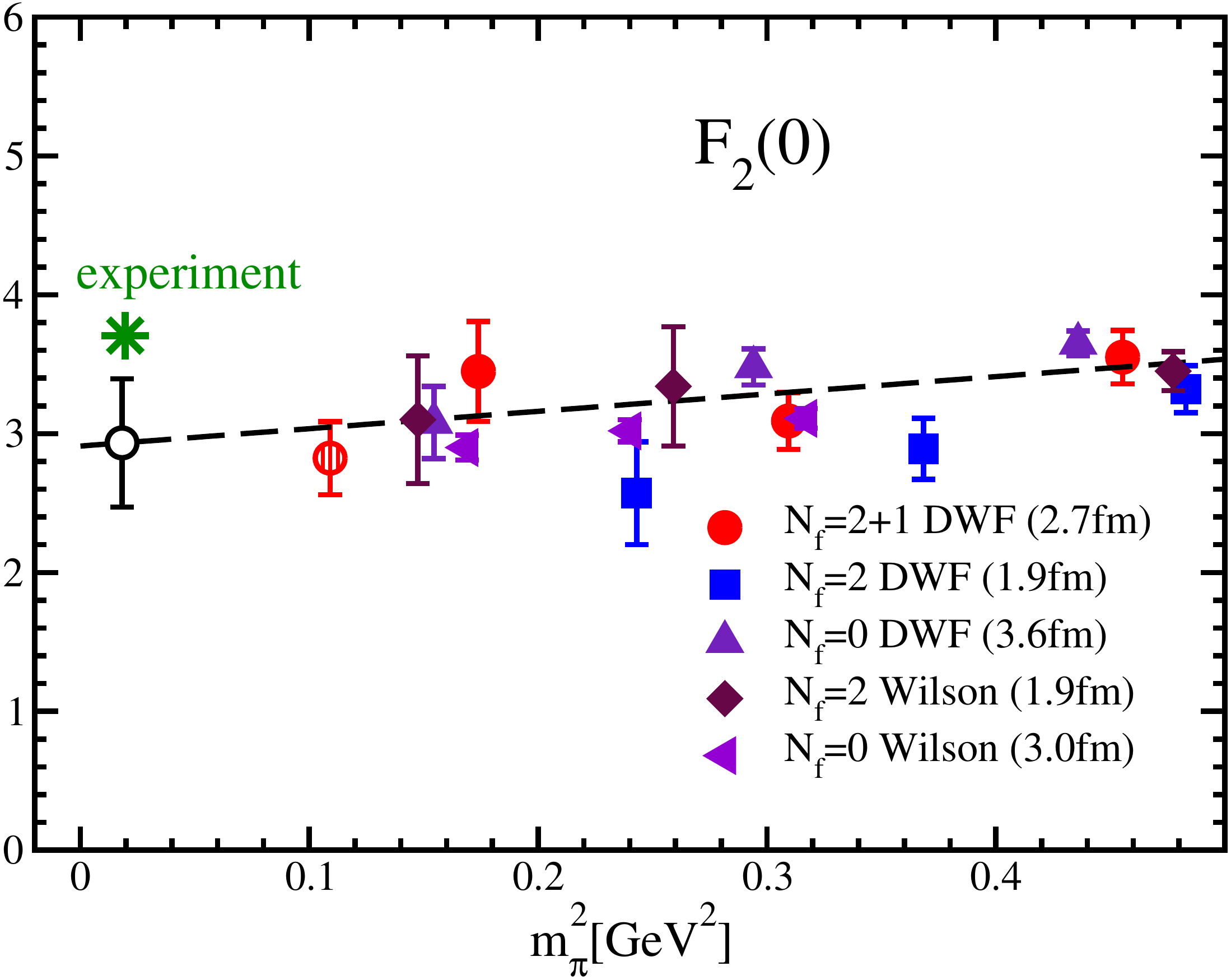}
\caption{Isovector vector current form factors.  Left: Dirac form factor mean square radius.  Center: Pauli form factor mean square radius.  Right: isovector anomalous magnetic moment.
The former two show no sign of expected singularity at \(m_\pi=0\), and significantly undershoot experiments.
The latter seems in rough agreement with the experiment.
}
\label{fig:ff}
\end{figure}

In contrast to the axialvector-current form factors, the vector-current ones do not show dependence on the lattice size even at the lightest pion mass of about 330 MeV.
Mean squared radii of Dirac and Pauli form factors are obtained (see Fig.\ \ref{fig:ff}) without chirally expected singular behavior in \(m_\pi^2\), and undershoot the experiments.
On the other hand the anomalous magnetic moment is in rough agreement with the experiment (Fig.\ \ref{fig:ff}).
These observations are confirmed by a LHP study using our 2.2-GeV ensembles \cite{Lin:2010ne}.

\begin{figure}[b]
\includegraphics[width=.5\linewidth,clip]{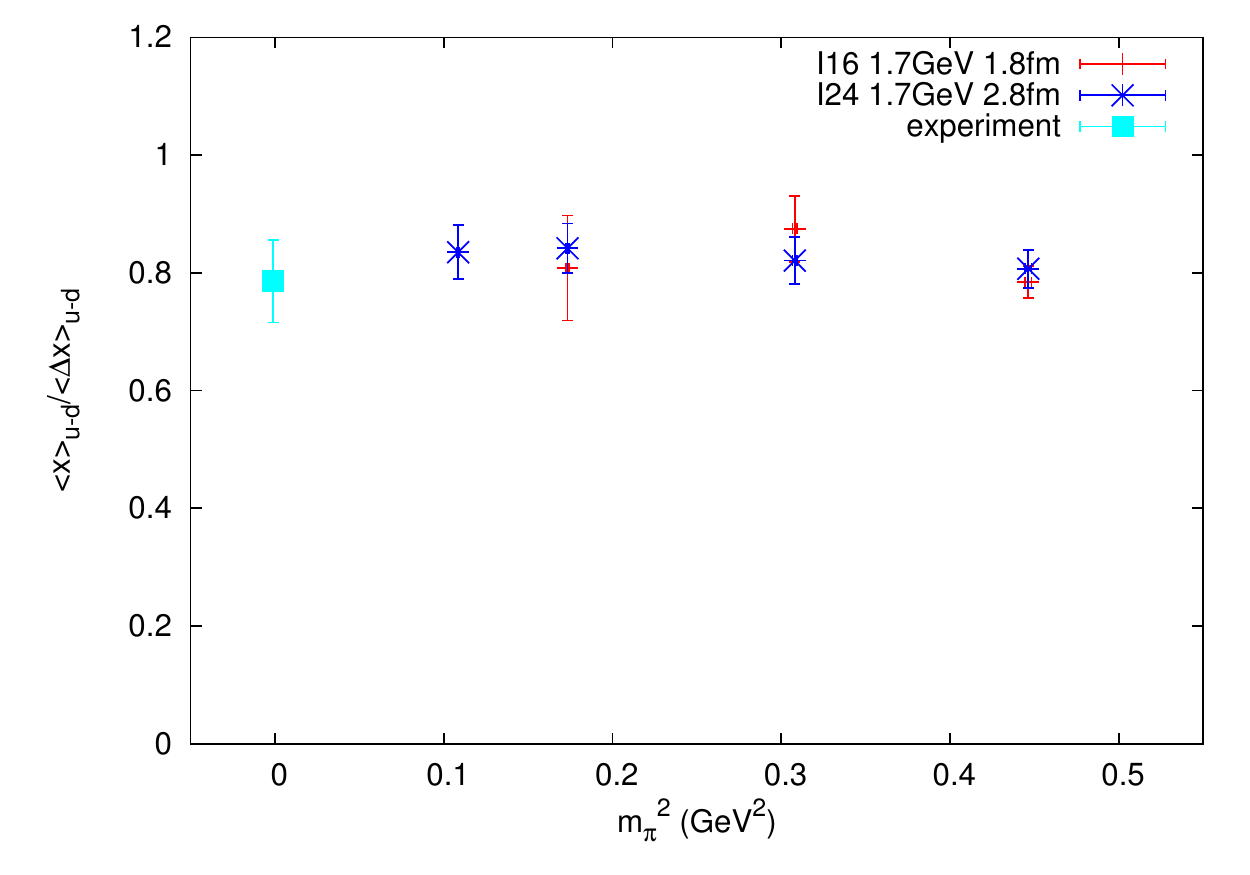}
\caption{
Naturally renormalized ratio, \(\langle x \rangle_{u-d}/\langle x \rangle_{\Delta u - \Delta d}\), of the isovector quark momentum and helicity fractions, does not show dependence on mass nor lattice size and agree with experiment.
}
\label{fig:sfratio}
\end{figure}
\begin{figure}[tb]
\includegraphics[width=.5\linewidth,clip]{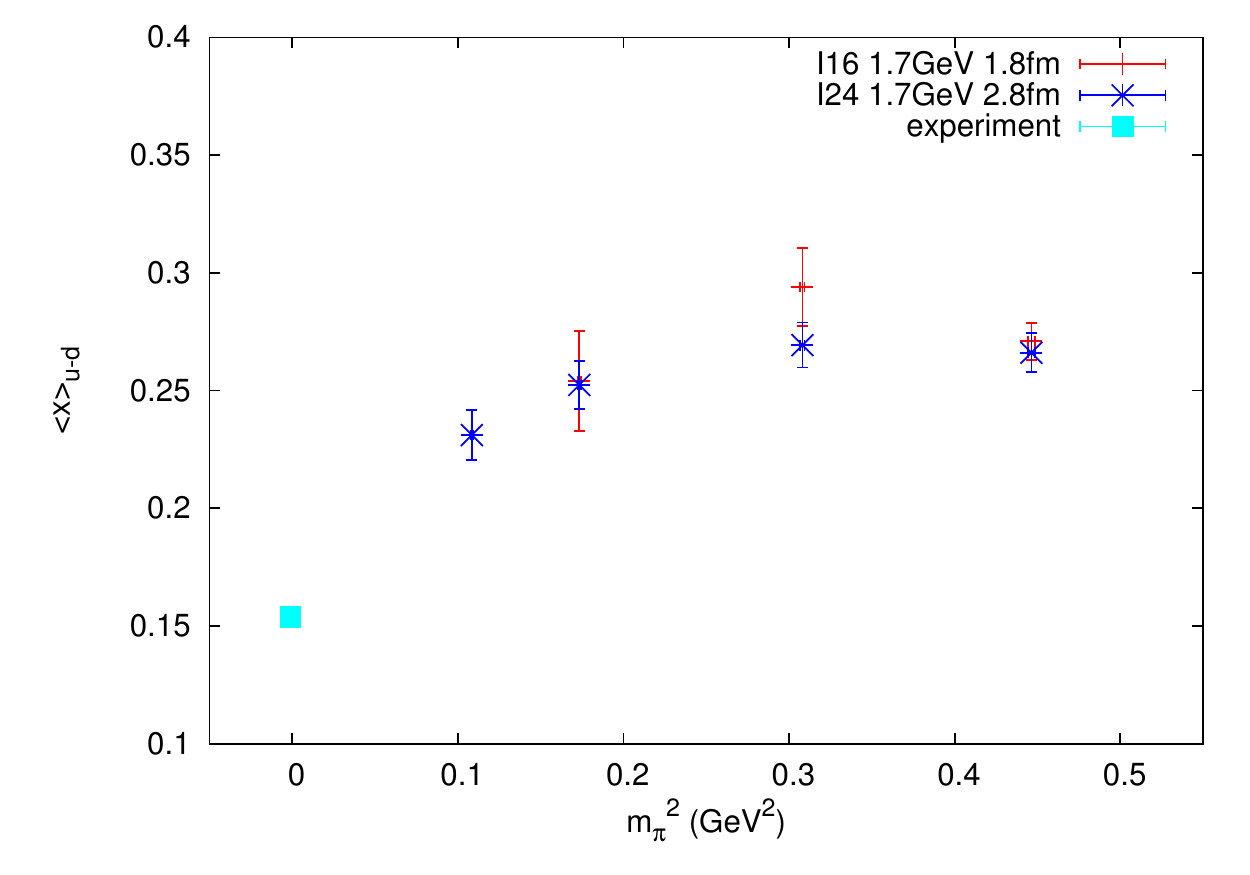}
\includegraphics[width=.5\linewidth,clip]{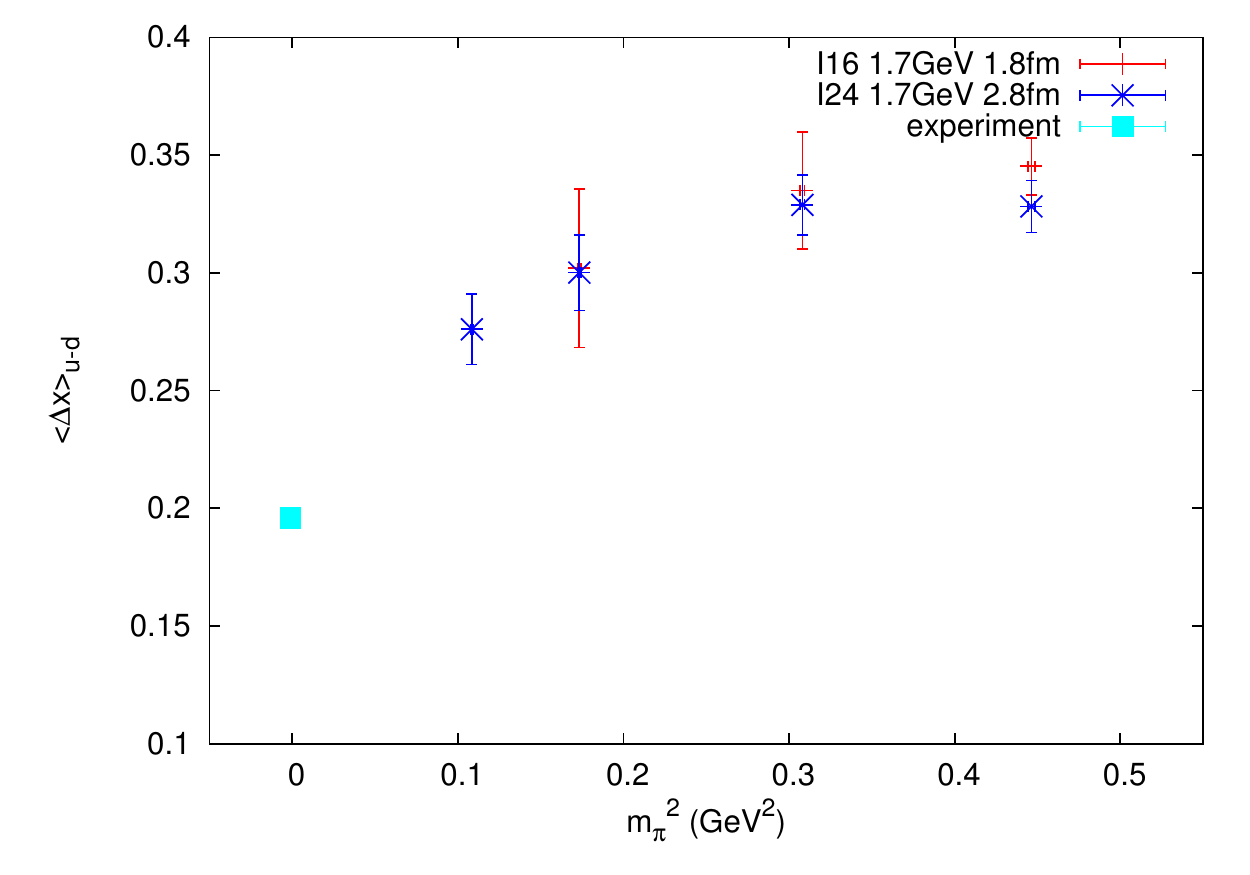}
\caption{
Isovector quark momentum fraction, \(\langle x \rangle_{u-d}\) (left), and helicity fraction, \(\langle x \rangle_{\Delta u - \Delta d}\) (right).
Non-perturbatively renormzlized and run to \(\overline{\rm MS}\) 2 GeV, \(Z^{\overline{\rm MS}({\rm 2 GeV})}=1.15(3)\),  and 1.15(4), respectively.
Neither shows volume dependence between larger, \(L=2.7\) fm, and smaller, 1.8 fm, lattices.
Both show interesting trend down to the experiment at the lightest mass.
}
\label{fig:sf}
\end{figure}

In low moments of isovector structure functions, such as quark momentum fraction, \(\langle x \rangle_{u-d}\), or helicity fraction, \(\langle x \rangle_{\Delta u - \Delta d}\), no dependence on \(m_\pi L\) is seen either.
In a naturally renormalized ratio, \(\langle x \rangle_{u-d}/\langle x \rangle_{\Delta u - \Delta d}\), of the two fractions, no dependence on the mass, \(m_q \propto m_\pi^2\), is seen either and the agreement with experiment is excellent (see Fig.\ \ref{fig:sfratio}).
No dependence on \(L\) is seen in respective fractions either (Fig.\ \ref{fig:sf}).
Both show interesting trending down toward the experiment, motivating us to calculate these quantities at lower pion mass.

\section{Status at 1.4 GeV}

Our studies of (2+1)-flavor dynamical DWF lattice QCD in \((\sim{\rm 3 fm})^3\) spatial box and down to about 300 MeV pion mass \cite{Aoki:2010dy} clearly point to the need of numerical calculations at yet lighter pion mass, both for combined chiral and continuum limit study in the meson sector and for better understanding of chiral behavior in the baryon sector.
Thus RBC and UKQCD collaborations started to generate a new set of ensembles with pion mass about 180 and 250 MeV \cite{Mawhinney:LAT2010}.
As the lower pion mass demands larger lattice spatial extent, the new ensembles are generated with about 4.6 fm spatial extent that translates to the scaling parameter \(m_\pi L\) of above 4 for the lighter mass and almost 6 for the heavier.
This is made possible by a newly developed gauge action with an appropriate addition of dislocation-suppressing-determinant-ratio (DSDR) term \cite{Vranas:1999rz,Vranas:2006zk,
Renfrew:2009wu}.
We have about 1900 hybrid Monte Carlo time units for the heavier ensemble and 1200 for the lighter, of which first 600 and 500 respectively are discarded for thermalization.
We analyze every eight time unit with four evenly separated nucleon sources each.

We use Gaussian smearing \cite{Alexandrou:1992ti, Berruto:2005hg} for nucleon source to optimize the overlap with the ground state and compared the cases for widths 4 and 6 lattice units.
We found width 6 is better for both pion mass \cite{ohta:2010sr}.
We quote preliminary nucleon mass estimates of 0.721(13) and 0.763(10) lattice units which correspond to about 0.98 and 1.05 GeV with another preliminary estimate for the lattice cut off of 1.368(7) GeV.
\begin{figure}[tb]
\includegraphics[width=.7\linewidth,clip]{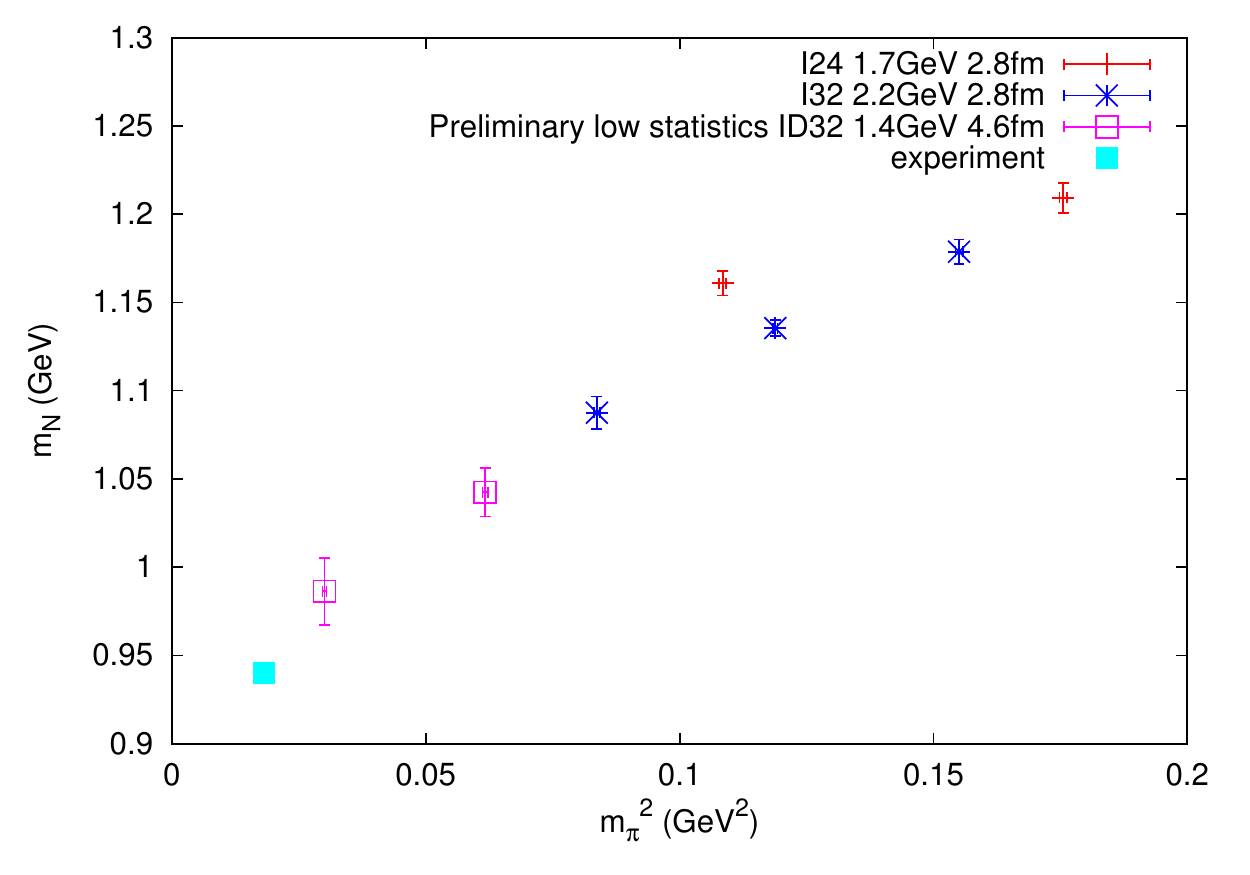}
\caption{
Nucleon mass from the RBC and UKQCD (2+1)-flavor dynamical DWF ensembles plotted against corresponding pion mass squared.  I24 and I32 are from ensembles with simple Iwasaki gauge action while ID32 are with new Iwasaki+DSDR gauge action.
}
\label{fig:summary}
\end{figure}
As are summarized in Fig.\ \ref{fig:summary}, these ensembles are filling the gap toward the physical point.
We look forward to reporting nucleon form factors and low moments of structure functions soon.


\begin{theacknowledgments}
I thank RBC and UKQCD Collaborations, especially Meifeng Lin, Yasumichi Aoki, Tom Blum, Chris Dawson, Taku Izubuchi, Chulwoo Jung, Shoichi Sasaki and Takeshi Yamazaki.
RIKEN, BNL, the U.S.\ DOE, University of Edinburgh, and the U.K.\  PPARC provided  facilities essential for the completion of this work.
The I+DSDR ensembles are being generated at ANL Leadership Class Facility (ALCF.)
The nucleon two- and three-point correlators are being calculated at RIKEN Integrated Cluster of Clusters (RICC.)
\end{theacknowledgments}

\bibliographystyle{aipproc}   

\bibliography{TQCD2010}

\IfFileExists{\jobname.bbl}{}
 {\typeout{}
  \typeout{******************************************}
  \typeout{** Please run "bibtex \jobname" to optain}
  \typeout{** the bibliography and then re-run LaTeX}
  \typeout{** twice to fix the references!}
  \typeout{******************************************}
  \typeout{}
 }

\end{document}